\documentclass[dvips]{article}
\usepackage{amsmath}
\usepackage{amssymb}
\usepackage{icrctc07}

\title{Tau neutrino search with the MAGIC telescope}
\shorttitle{Tau neutrino search with MAGIC}

\authors{M.\,Gaug$^{1}$, C.\,Hsu$^{2}$, J.\,K.\,Becker$^{3}$, A.\,Biland$^{4}$, M.\,Mariotti$^{1}$, W.\,Rhode$^{3}$, M.\,Teshima$^{2}$}
\shortauthors{Gaug, Hsu and Becker et al.}
\afiliations{$^1$Instituto de Astrof\'isica de Canarias, C/ Via L\'actea s/n, 38205 La Laguna, Tenerife, Spain\\ 
  $^2$Max-Planck-Institut f\"ur Physik, F\"ohringer Ring 6, 80805 M\"unchen, Germany\\
  $^3$Institut f\"ur Physik, Universit\"at Dortmund, 44221 Dortmund, Germany\\
  $^4$ETH Z\"urich, Institute for Particle Physics, 8093 Zurich, Switzerland}
\email{gaug@pd.infn.it, cchsu@mppmu.mpg.de, julia.becker@udo.edu}

\abstract{The MAGIC telescope located on the Roque de los Muchachos on the
Canary Island La Palma at a height of 2200\,m a.s.l. is able to
point to the sea. This permits a search for air shower signatures
induced by particles coming out of the Earth. 
An analytical
approximation results in $\nu_\tau$ effective areas from $\sim 10^3$ m$^2$ (at
100~TeV) to 10$^5$ m$^2$ (at 1~EeV) for an observation angle of about 
1$^\circ$ below the horizon, rapidly diminishing with 
further inclination. Taking into account
the huge effective area, this configuration was investigated for its suitability to search
for ultra-high energy (UHE) tau-neutrino signatures.
The outcome of simulations for tau-neutrino signatures will be presented, models for astrophysical
neutrino sources reviewed, and estimated event rates in MAGIC are shown.
}

\begin{document}
\maketitle
\section{Introduction}

\begin{figure}[h!]
\centering
\noindent
\includegraphics [width=0.5\textwidth]{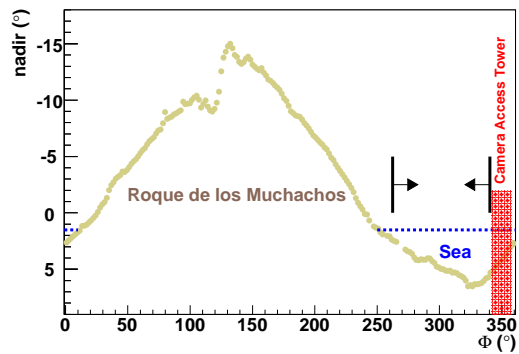}
\caption{Horizon seen from the MAGIC Telescope. The region $\Phi=[262^\circ,340^\circ]$ can 
be used to point toward the Sea.}\label{fig1}
\end{figure}

Although optimized to detect electromagnetic air showers
produced by gamma ray primaries, the MAGIC telescope~\cite{magic} (2200\,m a.s.l., 
28.45$^\circ$N, 17.54$^\circ$W) 
is also sensitive to hadronic showers. 
When the telescope turns down, this background 
diminishes until almost vanishing at the horizon. 
The telescope can look down to a maximum of about $10^\circ$ below the 
horizontal plane, the Sea is visible in an azimuthal range covering about 80$^\circ$ 
(see fig.~\ref{fig1}). Only 
a small light contamination from continuous scattered star light 
and from scattered Cherenkov light initiated by air showers will then be observed. 

We investigated the possible response of the MAGIC telescope 
to upwards moving showers initiated by UHE $\tau$-particles 
originating from a $\nu_\tau$ collision with the Sea or underneath rock. 
\par
The $\nu_\tau$ channel has several advantages with respect to the $\nu_e$
 or $\nu_\mu$ channel. First, the majority of the possible $\tau$ 
decay modes leads to an (observable) air shower or a combination of showers.
 Only 17.4\% decays to a muon and neutrinos, considered to be unobservable for the effective areas of 
interest here. Moreover, the boosted $\tau$ lifetime ranges from $\sim$50~m at 1~PeV to almost hundred kilometer at 
EeV energies~\cite{dutta,tseng}, only slightly affected by energy losses in matter. 

The production of UHE neutrinos in astrophysical shocks is expected in
the case of hadronic models, where accelerated protons interact with photons via the 
Delta resonance.
This process leads to the coincident production of neutrinos and TeV photons,
since charged pions have neutrinos as decay products while $\pi^0$ mesons
decay into two photons.
Alternatively, proton-proton interactions lead to a similar output of neutrinos.
The production ratio of the different flavors of neutrinos is
$(\nu_e:\nu_{\mu}:\nu_{\tau})_\mathrm{source}=(1:2:0)$.
Neutrino oscillations will equalize their rates, so that 
$\nu_e$, $\nu_\mu$ and $\nu_\tau$ should be detected in
equal fractions even though they are not emitted as such.


\section{Expected effective $\tau$-neutrino areas}

In order to estimate the effective areas of an Imaging Air Cherenkov telescope looking down to the Sea from 
2200\,m altitude, a small simulation was written for the case of a $\nu_\tau$ entering 
the Earth, creating a $\tau$-particle there (or in the Sea) and exiting the Sea towards the telescope. 
Also the case of a two-, three- and four-fold interaction-decay sequence was simulated. 
The $\tau$ subsequently decays and gives rise to an energetic air shower in 82.6\% of the cases. 
Only mean interaction lengths were simulated (in a similar way as found in~\cite{feng,aramo}), 
and we assume a shower opening angle of 1.4$^\circ$ and an effective 
trigger area of 2$^\circ$ of the MAGIC camera. Above $2\cdot 10^8$~GeV shower energy, the full 
2$^\circ$ field-of-view was taken as effective shower opening angle, due to the contribution 
of the fluorescence light. The depth of the Sea was assumed to be constantly 3~km throughout 
the observed area, with standard rock underneath. 
The $\tau$-particle range was assumed to 
be equal to its life-time (up to an energy of 10$^8$~GeV), above this energy, 
energy losses in the parameterization of~\cite{dutta} (formula~16) were computed.
\par

\begin{figure}[h!]
\centering
\includegraphics[width=0.5\textwidth]{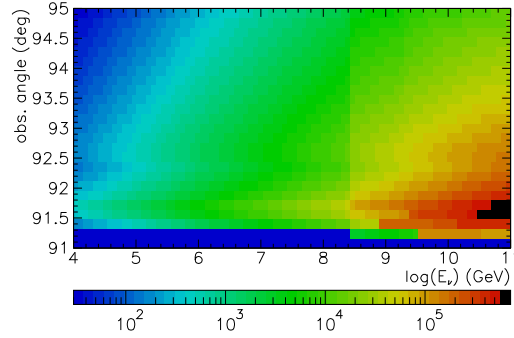}
\caption{Effective tau neutrino areas (in m$^2$).}
\label{fig7}
\end{figure}

Under these assumptions, we obtained energy and zenith angle dependent effective areas as 
shown in fig.~\ref{fig7}. One can see that basically a range of zenith angles around 91.5$^\circ$ to 
maximally 92.5$^\circ$ yields reasonable effective areas, reaching $5\cdot10^5$~m$^2$ at 
energies around 100~EeV. The traversed distance in Earth amounts then to about 100~km, an optimal value also 
obtained in the literature~\cite{bertou,fargion,hou,tseng}.

\section{Determination of the Energy Threshold}

When the telescope looks down to the Sea from a mountain of 2200\,m height, the horizon is 
seen under a zenith angle of $\sim$91.5$^\circ$,
at a distance of $\sim$165\,km.
Moving down to 94$^\circ$,
 the surface of the Sea is still $\sim$32\,km away. These numbers have to be compared with the 
typical position of a shower maximum at 10~km height, obtained when the telescope looks directly upwards 
and observes showers at the trigger energy threshold of ~50\,GeV.

\begin{figure}[h!]
\centering
\includegraphics[width=0.42\textwidth]{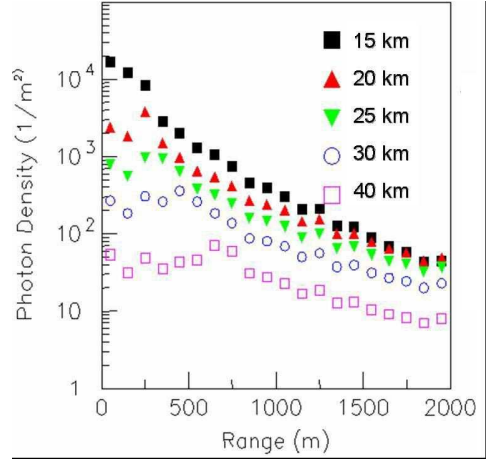}
\caption{Photon densities of Cherenkov light from a 1~PeV shower, observed at different 
distances.}
\label{fig8}
\end{figure}

Fig.~\ref{fig8} shows the simulated (observable) photon densities for an electromagnetic shower 
of 1~PeV injection energy, seen from various distances along the surface of the Earth. 
The MAGIC telescope has an effective mirror area of 236~m$^2$, with a photo-conversion efficiency 
of about 15\% at green to reddish wavelengths. 
Applying an analysis threshold of about 200 photo-electrons per image, 
a minimum Cherenkov photon density of about 7 photons/m$^2$ has to be required. For observations 
at 94$^\circ$ zenith angle, this limit translates directly into an energy threshold of about 40~TeV.
At 92$^\circ$ (91.7$^\circ$), the threshold becomes 300~(500)~TeV. 

This condition does not yet include the possible confusion with residual backgrounds from cosmic rays 
which will probably raise the thresholds further.

According to~\cite{fargion04}, muon bundles should penetrate to the telescope in two thirds of the cases, 
already below the threshold energies. However, we consider here the possibility to separate these signals 
from stray muons from cosmic rays extremely difficult, and do not predict any possibility to lower the
 thresholds even further. 

\section{Model Predictions}
Since the lower energy threshold for neutrino detection with MAGIC lies at
$E_{\min}\sim50$~TeV, the detection of galactic neutrinos can be
excluded, since sources like microquasars have typical maximum energies of
around $E_{\max}\sim 100$~TeV, see e.g.~\cite{torres}. In this
section, we review extragalactic sources suitable for observation with the
MAGIC telescope. 
\begin{figure}[h!]
\centering
\includegraphics[width=0.5\textwidth]{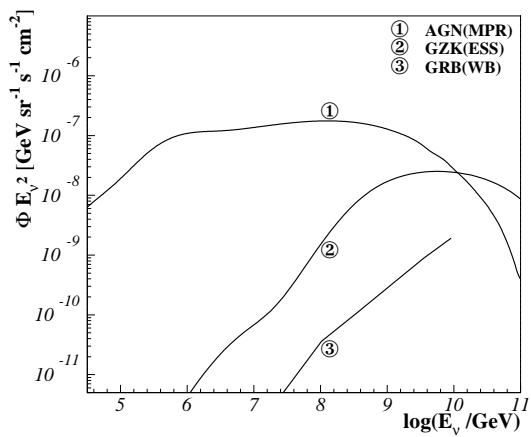}
\caption{Isotropic neutrino flux prediction from extragalactic sources. With
   MAGIC, the observation of single sources is done in the case of GRBs and
   AGNs. With the observation of the strongest sources, a higher flux than
   average can be expexted.}
\label{fig3}
\end{figure}
\subsection{Neutrinos from AGNs}
The observation of TeV photons from Active Galactic Nuclei are one indicator that also neutrinos
can originate from these sources, assuming that the TeV photons come from
$\pi^{0}$-decays. In the case of optically thick sources, keV-GeV emission can
also point to neutrino emission, if the UHE photon emission is a
cascaded TeV signal. Prediction (1) in fig.~\ref{fig3}, labeled {\it AGN(MPR)}, shows the maximum contribution from
GeV blazars as derived in~\cite{mpr}. There is a significant contribution up
to energies of $E\sim 10^{11}$~GeV, making it possible to test the model by
observing the most luminous GeV sources with MAGIC. In particular,
1ES~1959+650 is a good candidate for neutrino emission. In 2004, an orphan
flare, i.e.~TeV emission with no X-ray counterpart, has been detected from this source,
pointing to a partially hadronic origin of the radiation~\cite{1es1959_orphan}. The
neutrino emission from such a flaring state has been calculated
in~\cite{halzen_hooper_1es}. This model will also be used for the calculation
of expected event rates.
\subsection{GZK neutrinos}
UHE protons originating from extragalactic sources interact
effectively with the Cosmic Microwave Background via the Delta resonance above
energies of $\sim 10^{19.5}$~eV~\cite{greisen}. This process yields a neutrino
flux at energies of around $\sim 10^{8}-10^{11}$~GeV. Such a flux is
presumably present permanently and isotropically.
As an estimate, the prediction
from~\cite{ess} is shown as model (2) in fig.~\ref{fig3}, labeled {\it GZK(ESS)}.
\subsection{GRB afterglow neutrinos}
The prompt gamma-ray emission of Gamma Ray Bursts is typically followed by an
afterglow of X-ray, optical and radio radiation, which can be associated to an
UHE neutrino flux as pointed out in \cite{afterglow_nus}. Assuming
neutrino production via the Delta-resonance, the photon and proton energies
are correlated as $E_{\gamma}\cdot E_{p}\sim 0.2$~GeV$^2\cdot \Gamma^2$ with
the boost factor of the shock $\Gamma>100$. With a
typical afterglow photon energy of $\sim 100$~eV, the produced neutrinos will
have energies around $10^{9}$~GeV. Thus, such a flux is well-suited to be
investigated with MAGIC. The average neutrino spectrum from GRB afterglows is
shown as model (3) in fig.~\ref{fig3}, labeled {\it GRB(WB)}. However, GRB
spectra can in some cases deviate from the mean spectrum. 

\section{Results}

\begin{figure}[h]
\centering
\includegraphics[width=0.5\textwidth]{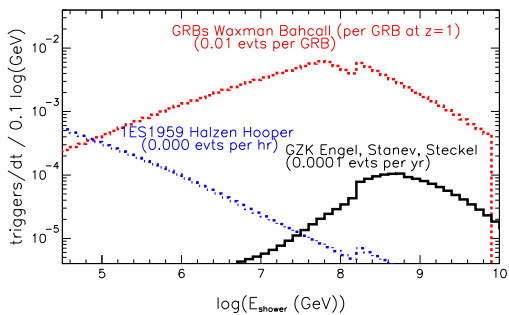}
\caption{Expected visible shower event rates from investigated neutrino models}
\label{fig6}
\end{figure}

Based on the above models, we estimated typical observable event rates from the 
AGN 1ES1959~\cite{halzen_hooper_1es}, a GRB located at a redshift of $z=1$~\cite{afterglow_nus} 
and the observation of the diffuse GZK neutrinos~\cite{ess}, displayed in figure~\ref{fig6}. 
Only the observation of a GRB at small distance seems  to give event rates that are potentially observable. 
Several strong bursts per year can be expected, possibly with a neutrino flux much higher than assumed in 
the figure.
These bursts can be observed by
the MAGIC telescope using an online trigger from GRB satellites once the GRB
occurs in the field of view of MAGIC neutrino observations.

\section{Acknowledgements}
The construction of the MAGIC Telescope was mainly made possible by the support of the German BMBF and MPG, 
the Italian INFN, and the Spanish CICYT, to whom goes our grateful acknowledgment. 
We would also like to thank the IAC for the excellent working conditions at the 
Observatorio del Roque de los Muchachos in La Palma. This work was further supported 
by ETH Research Grant TH 34/04 3 and the Polish MNiI Grant 1P03D01028.

\end{document}